# Non-modified Thermally-derived Onion-like Carbon As Electrocatalyst for [VO]²⁺/[VO₂]⁺ Redox Flow Battery

**Young-Jin Ko[†,‡], Jung-Min Cho[†,‡], Doo Seok Jeong[†], Inho Kim[†], Heon-Jin Choi[‡] and Wook-Seong Lee*,[†]**

[†] Electronic Materials Center, Korea Institute of Science and Technology, Seongbuk-gu, Seoul 136-791, Republic of Korea, [‡] School of Advanced Materials Science and Engineering, Yonsei University, Seoul, 120-749, Republic of Korea

**ABSTRACT**

We report the nanodiamond-derived onion-like carbon successfully applied as an electrocatalyst for [VO]²⁺/[VO₂]⁺ redox flow battery, as drop-coated (in the as-synthesized state) on glassy carbon or carbon felt electrodes. We show that its reversibility and catalytic activity in its as-synthesized state was comparable to some of the best data in the literature which employed surface modifications. We clarified the origin of such excellent performances by physical/electrochemical analyses.

Vanadium redox-flow battery (VRFB) is the next-generation large-scale electrochemical energy-storage system,[1, 2] which benefits from low cross-over (owing to using identical electrolytes (vanadium ions and vanadium oxide ions) on both electrodes),[3] low electrode damage (owing to the electrode-surface-confined redox reaction of the electrolytes) and consequently high cycling stability.[4] Nevertheless, the activity and the reversibility of the most commonly adopted electrode material, i.e. the carbon felt or carbon papers,[5] are poor due to the multi-step [VO]²⁺/[VO₂]⁺ coupled reaction at the cathode.[6] The usual approach to this issue is the catalytic agent dispersion or the direct surface modification. The catalytic agents include 1) the noble metal nanoparticles,[7-9] 2) oxygen-functionalized nanocarbon,[10-16] 3) graphene modified by heteroatom (mostly nitrogen) doping.[6, 17, 18] The direct modification of the carbon electrode is carried by the heteroatom doping or by the oxygen-functinalizations.[6, 14] However, such approaches suffer the risk of hazardous chemicals (heteroatom doping or oxidation),[19-21] and inconvenience of complex processing, or high-cost (noble metal nanoparticles). Alternative approach has been in strong demand, but relevant studies were rare.

Recently, Park *et al.* reported an approach that resolved such issue, based upon the electrochemically-active *inherent* defect sites on the CNF, hybridized with the good electrical conductivity of CNT, co-grown by CVD on the carbon felt.[22] On the other hand, we recently reported an *inherent* defect generation mechanism (based upon the diamond core volume expansion accompanying the diamond-to-graphite phase transformation) in nanodiamond-derived onion-like carbon (OLC) and consequent strong electrochemical activation, equivalent to that enabled by the doping and post-treatments, in dopamine sensing.[23] The OLC was prepared by simple vacuum annealing of the cheap detonation nanodiamond, free from additional catalytic agents or post-treatments, well drop-coated on carbon felt. Such properties strongly suggested that it might as well be applied as an electrocatalyst for [VO]²⁺/[VO₂]⁺ redox flow battery, since the latter also involves electron transfer steps[12] as the former did.[23] The adopted simple thermal synthesis process might also provide a facile route for controlling the *types* and densities of the generated inherent defects, which might be important for optimizing the intrinsic redox properties, as well as for studying the relevant reaction path as will be shown here. So far such issues were not clarified yet, even for the cases of the previously reported electrocatalytic agents,[6, 12] let alone those concerning the OLC. Here we investigated this issue, by studying the temperature-dependent inherent defect evolution in the nanodiamond-derived OLC and the consequent electrocatalytic activity evolution in [VO]²⁺/[VO₂]⁺ redox flow battery.

**Experimental Method**

*Materials*: All materials were purchased from Sigma Aldrich Co. All aqueous solutions were prepared with deionized water (DI water, ionic resistivity = 18.2M /cm). The [VO]²⁺ solutions were prepared in 3M sulfuric acid solution.

*Synthesis of OLCs*: The OLC powders were synthesized by vacuum annealing of the nanodiamond powder in the tube furnace. The furnace was evacuated with rotary pump to 10⁻³torr with the annealing time fixed at 1 hour. The annealing temperatures were varied as 1000°C, 1200°C, 1400°C, 1600°C, 1800°C and 2000°C; the samples were accordingly denoted as OLC-1000, OLC-1200, OLC-1400, OLC-1600, OLC-1800 and OLC-2000. The annealed samples were furnace-cooled after the annealing process.

*Preparation of working electrodes*: The glassy carbon electrode (GCE) of 0.247cm² area was polished with alumina abrasive powders prior to drop coating. It was subsequently rinsed with aqueous solution of Isopropyl alcohol (IPA) and was dried at room temperature. Five milligrams of synthesized OLC powder were ultrasonically blended with IPA (0.5mL), DI water (1.5mL) and 5wt% nafion solution (50μL). Twenty microliters of catalyst ink was loaded onto the pre-treated GCE and was dried at room temperature. The mass loading was 0.0195mg/cm².

Prior to half-cell test of Carbon felt (CF, PAN CF-20-3, Nippon carbon), catalyst inks were prepared by mixing ten milligrams of synthesized OLC powder in mixture of IPA (0.9mL) and Nafion solution (5wt%, 100μL). The 5mg/cm² of suspension was dropped onto the CF and dried at 60°C for 12hr.

*Electrochemical Measurement*: A conventional three-electrode system was used throughout all the electrochemical measurements. A platinum wire and the Ag/AgCl electrode were used as the counter electrode and the reference electrode, respectively. The electrochemical impedance spectroscopy was carried out at open circuit voltage (OCV) from 10⁵Hz down to 10⁻²Hz with 10mV AC amplitude.

*Structural and Physical characterization of OLC*: The microstructure of OLCs was analyzed by high-resolution transmission electron microscopy (HR-TEM, FEI Co., Titan 300kV). The defect density and strain of OLCs was characterized by visible-Raman spectroscopy (Renishaw Co., inVia Raman spectroscopy, Beam Source: 532nm Nd:YAG laser). Since the visible-Raman is 50~230 times more sensitive to sp²-carbon than sp3-carbon, the sp³-carbon signal was negligible. Curve fitting for the measured spectra was carried out according to the previous report.[24]

The density of state (DOS) was analyzed by ultraviolet photoemission spectroscopy (UPS, Ulvac Co., PHI 5000 Versaprobe). A He(I) emission lamp (photon energy: 21.2eV, penetration depth: few Å) was used as beam source. The spectral resolution was 0.01eV and the electron take-off angle was 90°. Gold was used as the reference sample. The oxygen ratio of OLCs was characterized by X-ray photoemission spectroscopy (XPS, Ulvac Co., PHI 5000 Versaprobe).

The sp²-sp³ ratio of OLCs were analyzed by near edge x-ray absorption fine structure spectroscopy (NEXAFS, resolution: 0.1 eV, total electron yield mode, at the 10D KIST beam line of the Pohang Accelerator Laboratory in Republic of Korea). The sp²-sp³ ratio was calculated from the obtained NEXAFS spectra employing the technique of the previous report.[25]

The dangling bond densities of OLCs were measured by electron-spin resonance (ESR, JEOL Co., JES-FA100). Samples of 5~10 milligrams were put in the quartz tube and were placed in the cavity. The copper sulfate pentahydrate was used as reference sample (1 spin/molecule).

Dispersion of OLCs in water was carried out in order to visually confirm their hydrophobicity. Two milligrams of OLCs were wetted with 5mL DI water and dispersed in sonication bath. After the sonication, the mixtures were manually agitated and put to rest for 24 hours; subsequently they were visually inspected.

## Results and Discussions

Figure 1 shows the HR-TEM images of the OLCs derived from detonation nanodiamond annealed at various temperatures. The number of the graphitic shells increased with temperature at the expense of the diamond core, which agreed with previous report.[26] At 1600°C, the diamond core was completely consumed in the majority of the particles. While the graphitic shell was quasi-spherical upto 1600°C, the shell polygonization prevailed at 2000°C (Figure 1f). Figure S1 shows the particle size distributions at various annealing temperatures; the obtained averaged particles sizes were plotted versus the synthesis temperature in Figure S2: it rapidly increased upto 1200°C and the slop was reduced thereafter. The initial steep increase was attributed to the volume expansion arising from the diamond-to-graphite phase transformation,[23] while the slop reduction beyond 1200°C was attributed to the eventual consumption of the diamond core.[23]

The electrochemical activities were analysed for [VO]²⁺/[VO₂]⁺ redox reaction on the electrodes drop-coated with the aforementioned OLCs (glassy carbon electrode: Figure 2, carbon felt electrode: Figure 3). The peak currents for both types of electrode strongly increased with the annealing temperature upto 1800°C and dropped thereafter (Figures 2, 3). The parameters from the CV (peak potential difference: $\Delta E_p$, oxidation peak current: $I_P{}^{ox}$, peak current ratio: $I_{PA}/I_{PC}$) were summarized in Table S1 (glassy carbon) and Table S2 (Carbon Felt). Since the drop-coated OLC masses differed greatly in the two electrodes (Experimental), the redox probe concentrations were adjusted accordingly: 1M for the glassy carbon electrode and 0.1M for the carbon felt electrode. Table S3 compares the redox performance of the OLC-1800 with some of the best data in the literatures. It was remarkable that the peak current densities (Jp) of the OLC-1800 (neither oxygen-functionalized nor post-treated) were comparable to those of the various nitrogen-doped or oxygen-functionalized nano carbons, as well as those of the best nano-doped sample (CNF/CNT-700).Another important property, *i.e.* the reversibility is represented by two equivalent parameters: peak potential separation ($\Delta$Ep) and the redox peak currents ratio ($I_{PA}/I_{PC}$ ratio, see Table S2 and Figure S3). The reversibility is inversely proportional $\Delta$Ep or proportional to the proximity of the $I_{PA}/I_{PC}$ ratio to unity.[12] Figure S3 indicates that both of the reversibility parameters peaked at 1800°C in their temperature: in particular, the $I_{PA}/I_{PC}$ ratio was among the best, *i.e.* the closest to unity (Table S3).

While such redox performances of the OLC (peak currents and reversibility parameters) were remarkable, the origin of their temperature-dependent evolutions, which unanimously peaked at 1800°C, was intriguing. To study this issue, it was important first to clarify the rate-determining steps of the redox reaction. We analysed the CV to show that the present redox reaction was mixed-controlled, with electron transfer and the mass transfer coupled in series (Statement S1, Figures S4-S7 and Table S4-S5). Here it was important that many of the previous reports indicated the strong enhancement of the vanadium activity in the redox reaction by oxygen.[10-16] It prompted us first to confirm such possibility: the oxygen contents, as analysed by XPS (Figure S8), monotonically decreased with temperature, in contrast to the CV responses (anodic peak currents) which peaked at 1800°C (Figure 2, 3). It demonstrated the irrelevance of the oxygen contents and demanded further analyses. The temperature-dependences of the redox performances of the present OLC samples was reminiscent of the analogous observations of electrochemical activity enhancement in nanodiamond-derived OLC for dopamine sensing in 1000-1400°C range in our previous report.[23] It prompted us for analogous approach in the present study, but in an extended temperature range of 1000-2000°C not to miss the possible location of the optimum temperature at higher temperature domain, as well as with more extended series of analytical tools, and the first principle calculations to clarify the redox reaction steps, as will be shown below.

Figure 4a, b shows the Raman spectra of the OLC samples, which gives bulk-specific information concerning the lattice defect density.[27] The peaks were deconvoluted to five different bands, *i.e.* I, D, A, G, and D'.[24] The $I_D/I_G$ ratio derived from Figure 4a, b was summarized in Table S6 and was plotted in Figure 4c. The defect density represented by the $I_D/I_G$ ratio increased with temperature to peak at 1800°C and dropped again thereafter.[27] Figure 4d shows the temperature dependence of the $I_D/I_{D'}$ ratio, which represented the *types* of defects in the graphene.[28] It gradually increased with temperature from 3 to 7 and eventually past 14, which corresponds to differing defect types of grain boundary (3.6), vacancies (7.06) and sp³-type defects (13.9).[28] Hence the defect type pertaining to the OLC-1800, where the a number of properties shown so far peaked at, seemed to be the vacancies rather than grain boundaries or the sp³-carbon type defects. Figure S9 shows the corresponding Raman spectra in the 2D band region; they were blue-shifted with respect to the graphene, which indicated that the OLC shells were subjected to the tensile strain.[29, 30] The Figure S9c, d shows the 2D peak shift and the corresponding tensile strain calculated from the shift,[23] respectively; the strain peaked at 1400°C (Figure 5d).

Figure 5a shows the electron-spin resonance (ESR) spectra of the OLC samples; it gives information concerning the dangling bonds,[31] which might be attributed to the vacancy generation.[32, 33] The resonance peak became broader with temperature; deconvolution of the spectra gave two isolated peaks (the peak widths were plotted in Figure 5b), which indicated the overlap of a narrow peak and a broader peak.[34] Both peaks were attributed to the non-bonding π-electron:[34] while the narrow peak was attributed to the interference-free resonance, the broad peak was attributed to the interference from other orbitals.[35] The spin densities, which might be regarded as the dangling bond density,[36] was calculated from the narrow peak intensity (plotted in Figure 5c); it initially dropped from that of raw material (detonation nanodiamond) until 1400°C, probably due to temperature-dependent consumption of the diamond core in the OLC, and hence the corresponding removal of the initial dangling bonds therein.[31] Then it increased again to peak at 1800°C, which was obviously attributed to the continued defect (*i.e.* dangling bond) generation as formerly indicated by the $I_D/I_G$ ratio evolution (Figure 4c).

Figure 6a shows the UPS spectra of the OLC samples, which gives surface-specific information concerning the mid-gap DOS,[37] which controls the electron transfer kinetics in graphene[37] as well as in OLC.[23] Figure 6b shows the temperature-dependence of the DOS as normalized by the Au reference; it agreed with the temperature-dependence of the defect density obtained by the Raman spectra (Figure 4c). Such agreement was attributed to the mid-gap states generation by the lattice defect.[38] Figure S10 shows the spectra obtained by electrochemical impedance spectroscopy (EIS), which provided additional information concerning the electron transfer as well as the mass-transfer. For the EIS data fitting, the Randle circuit was adopted;[39] resulting parameters were summarized in Table S7. The semi-circle appearing at the high frequency domain represents the charge transfer resistance and the double layer capacitance,[39, 40] while the linear slope at the low frequency domain represents the mass transfer factor, *i.e.* the Warburg element.[39] The inverse of the charge transfer resistance, which represents the electron transfer kinetics,[39] was plotted in Figure S11. The temperature-dependence of Rct[-1] resembled those of $I_D/I_G$ ratio and DOS (Figures 4, 6) in that they initially increased and peaked at 1800°C. It indicated the dependence of the electron transfer kinetics on the defect-induced mid-gap states generations.

Figure S12 shows the sp²/sp³ carbon masses (contents) as analysed by the NEXAFS. We have shown that the sp²-carbon contents represented the extent of the diamond-to-graphite phase transformation (at the diamond core) of which the relevant volume expansion drove the defect-generating, forced tensile strain at the outermost OLC shell in 1000-1400°C region.[23] The sp²-carbon contents initially increased steeply with temperature upto 1200°C, while it kept increasing mildly upto 2000°C. It was in contrast to aforementioned properties (peak currents, defect density, DOS and the Rct[-1]) that unanimously peaked at 1800°C. It indicated that some new defect-healing mechanism became dominant over the aforementioned defect-generation mechanism, at 2000°C. A possible origin of such healing mechanism might be related to the shell polygonization occurring at 2000°C (Figure 1). The polygonization involves the transformation of the quasi-spherical shells to the shells with large portions of flattened facets, probably via the recrystallization process which is well-known to reduce the residual strain or stress, and hence the defect healing.[41, 42] It was further supported by the temperature-dependence of the interplanar spacing between the outermost shell and the 2nd shell, which were reported to be controlled by the vacancy density;[30, 31] we determined it from the HR-TEM pictures (Table S8, Figure S13). The spacing peaked at 1800°C and thereafter dropped back to the level of low temperature domain; it indicated the carbon structure changed from turbostratic carbon to graphite, which suggested the aforementioned recrystallization, and the consequent strain relaxation possibly accompanying the defect healing.[43] Such defect healing resuming at 2000°C was further supported by the report by Okotrub *et. al.*:[44] they reported the temperature-induced holed-structure in OLC, which might be regarded as vacancy clusters (using the x-ray emission spectroscopy) at 1900K, and its disappearance at higher temperature.[44] (see statement S2 for additional comments)

So far we have discussed the factors contributing to one of the rate-determining steps: the electron transfer. Now we discuss those for an additionally coupled rate-determining step, *i.e.* the mass transfer. Recall that the present Redox reaction is given as follows.

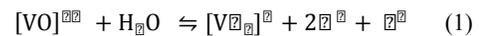

$$[VO]^{??} + H_? O \ \leftrightarrows \ [V?_?]^? + 2?^? + ?^?  \qquad (1)$$

There are two electrode-related factors relevant to the mass-transfer rate: the surface structure and the hydrophilicity.[6] The hydrophilicities of the OLC samples were compared by the water-dispersion test (Figure S14, visual inspection), which is widely adopted for the hydrophilicity test; in visual inspection, no meaningful difference was observed among the samples, except for the OLC-1000. It indicated that the mass transfer was determined by the surface structure rather than the hydrophilicity of the OLC. Among the various properties of the OLC shown so far, the most probable surface-structure-relevant factor was the dangling bond density represented by the ESR analyses (or, equivalently but somewhat less directly, the defect density represent by the $I_D/I_G$ ratio), for obvious reasons. Other properties, *e.g.*, DOS, is obviously relevant to the electron-transfer rather than the mass-transfer. Such proposition was strongly supported by the good agreements between the temperature dependence of the dangling bond density (Figure 5) with that of the peak currents (Table S2), in that they all peaked at 1800°C. The detailed reaction path through which such dangling bond evolution could affect the mass transfer, is in progress, via the first principle calculations.

## CONCLUSIONS

ND-derived OLC was successfully applied as an electrocatalyst for $[VO]^{2+}/[VO_2]^+$ redox flow battery. Reversibility and catalytic activity in the redox reaction was strongly dependent on the synthesis temperature; both properties peaked at 1800°C where they were comparable to some of the best data in the literature, although it was neither oxygen-functionalized nor post-treated. The redox reaction was mixed-controlled; the electron transfer step was controlled by the mid-gap density of states (DOS), induced by the mutually-competing defect generation /healing mechanisms unique to thermally-derived OLC. The mass transfer step was controlled by the chemical bonding of the vanadium oxide (as $[VO]^{2+}/[VO_2]^-$) to the carbon single dangling bonds of OLC. The dominant type of defect responsible for the optimized catalytic activity was the vacancy, rather than the grain boundary or sp³-type defects

*Acknowledgment.*

This work was supported by the institutional program grant (2E26370) from Korea Institute of Science and Technology. The authors are grateful to M.K. Cho (in Advanced Analysis Center, KIST) for the comments for HR-TEM characterization, to Dr.

K.-W. Chae at HoSeo University for his helps in the vacuum annealing process.

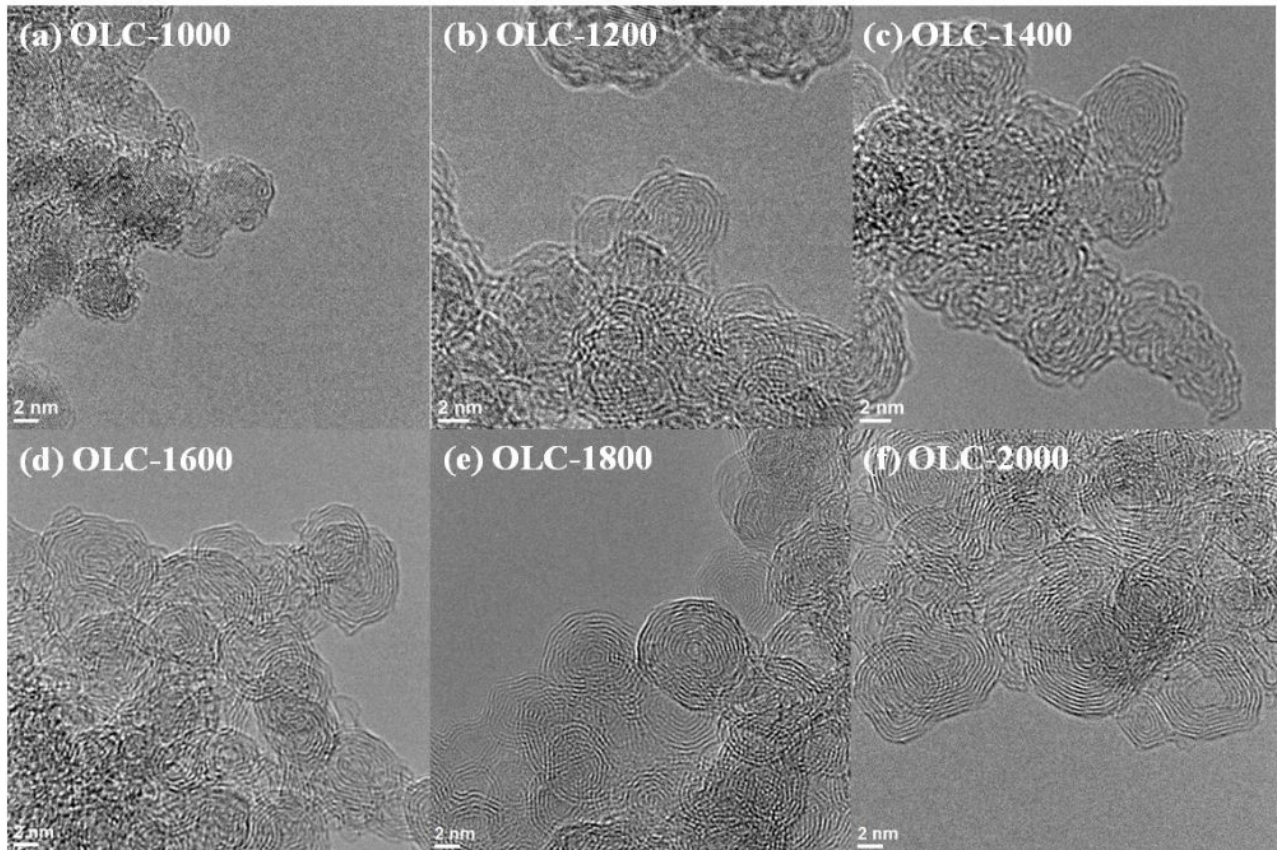

**Figure 1.** HR-TEM images of OLCs synthesized by annealing the detonation nanodiamond at temperatures of (a) 1000°C, (b) 1200°C, (c) 1400°C, (d) 1600°C, (e) 1800°C and (f) 2000°C, respectively.

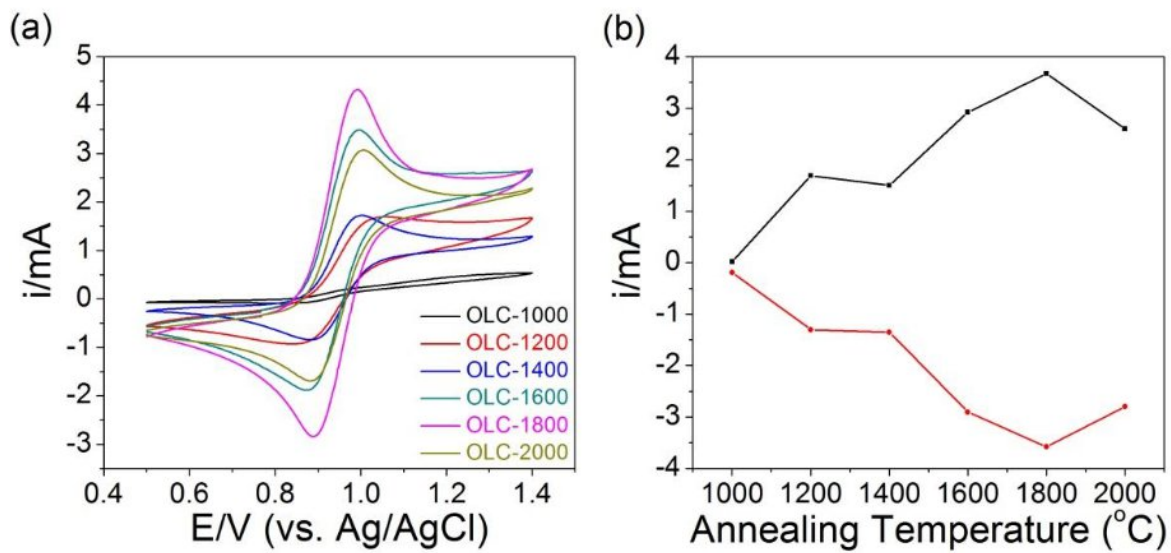

**Figure 2**. (a) CV responses from the OLC-modified glassy carbon electrodes in 1M of $VOSO_4$ + 3M $H_2SO_4$ solutions at scan rate of 5mV/s, (b) The anodic peak current *vs* annealing temperature profile (black line and scatter) and the cathodic peak current *vs* annealing temperature profile (red line and scatter) corresponding to Figure 2a.

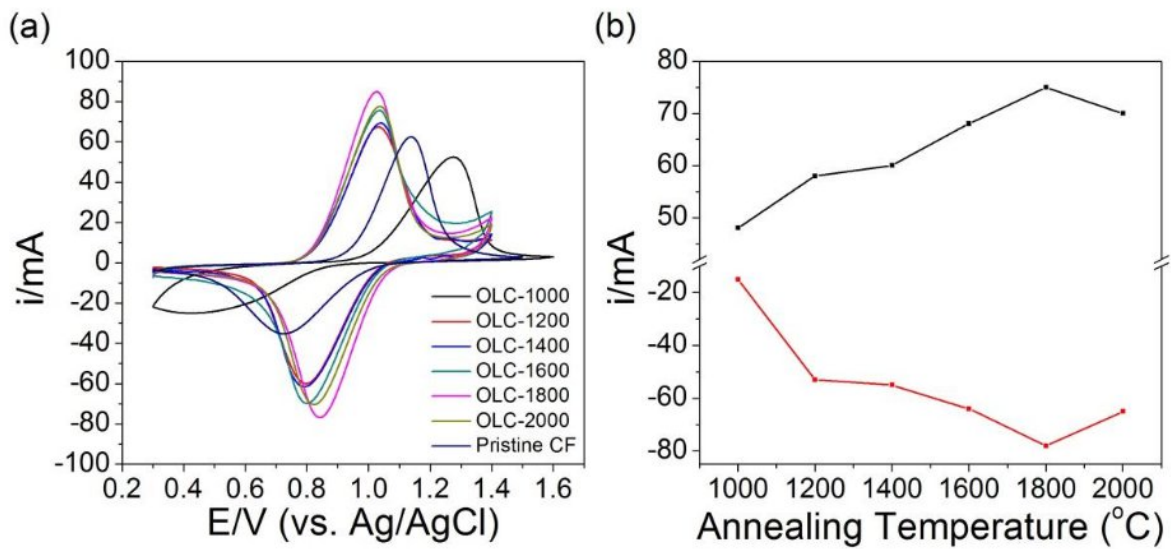

**Figure 3**. (a) CV responses from the OLC-modified CF electrodes in 0.1M of VOSO$_4$ + 3M H$_2$SO$_4$ solutions at scan rate of 5mV/s, (b) The anodic peak current *vs* annealing temperature profile (black line and scatter) and the cathodic peak current *vs* annealing temperature profile (red line and scatter) corresponding to Figure 3a.

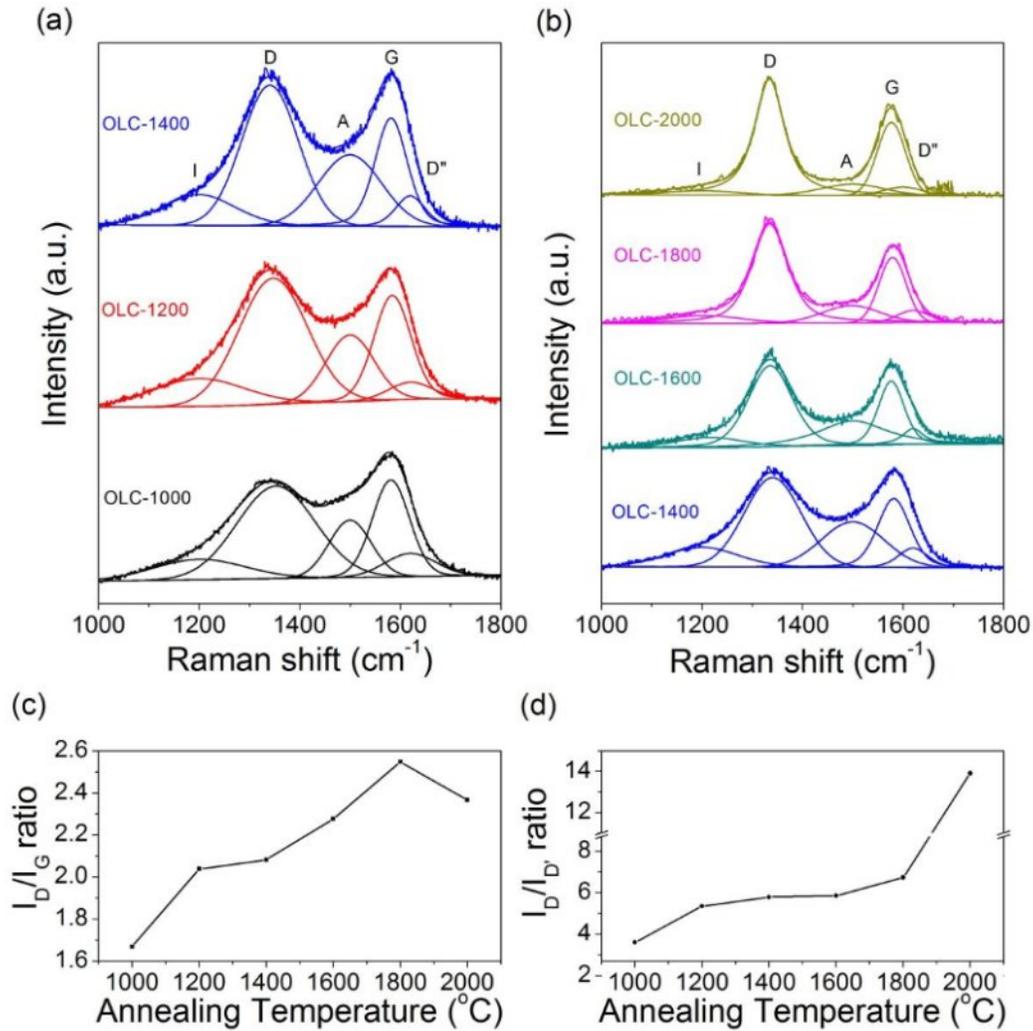

**Figure 4.** (a, b) Raman spectra of OLCs (from 1000cm[-1] to 1800cm[-1]) synthesized at various annealing temperatures; (c) $I_D/I_G$ ratio *vs* annealing temperature and (d) $I_D/I_{D'}$ ratio *vs* annealing temperature profile corresponding to Figure 4a, b

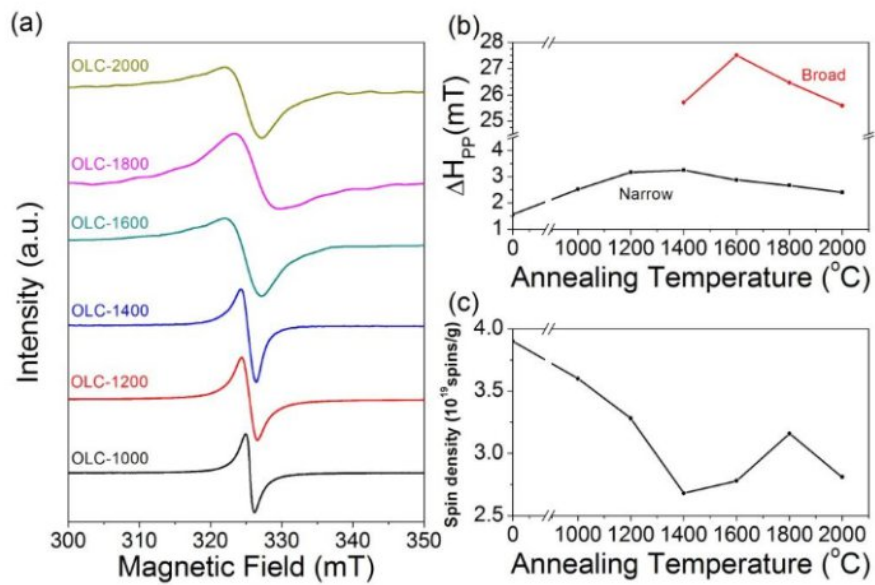

**Figure 5**. (a) The ESR spectra of the OLCs, (b) annealing temperature dependence of the ESR line widths ($\Delta H_{PP}$) and (c) the spin densities.

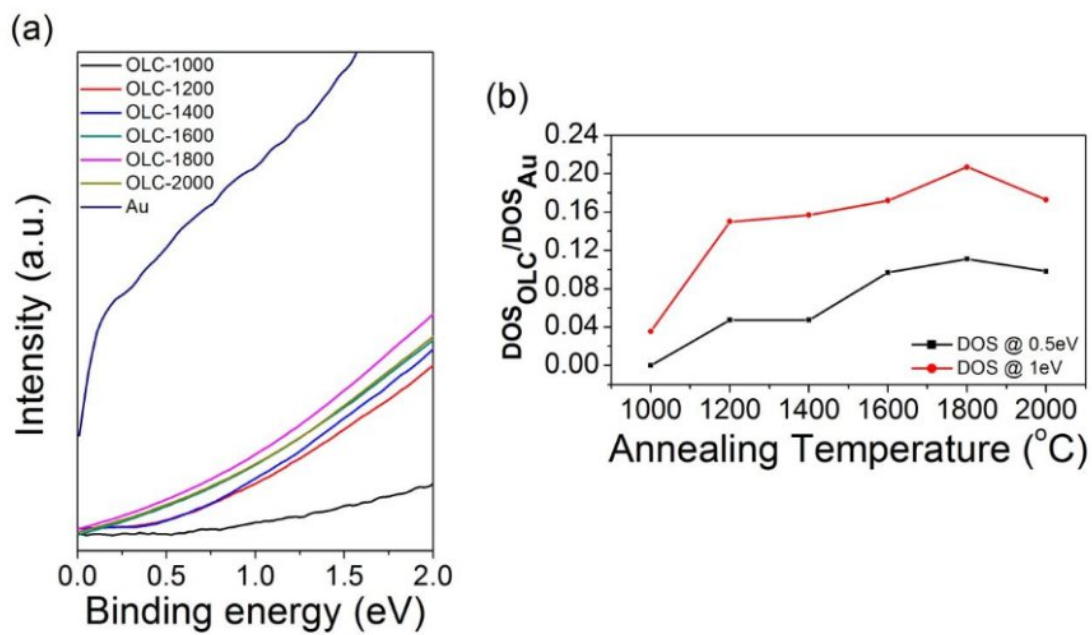

**Figure 6.** (a) UPS spectra of OLCs synthesized at various annealing temperatures and (b) The DOS of OLCs (normalized by the DOS of Au) *vs* annealing temperature profiles at various binding energy, corresponding to Figure 6a